# Planning for net zero by 2050, what HVAC system interventions will today's code minimum commercial buildings require?

**Patrick Pease, PE**
Member ASHRAE

**Jayati Chhabra**

**Zahra Zolfaghari**

**ABSTRACT**

*Heating, Ventilation and Air conditioning (HVAC) systems account for approximately 40% of the total energy used by buildings in the USA. To reduce this consumption States enforce minimum energy codes that currently range from strict (ASHRAE 90.1-2016) to relaxed (ASHRAE 90.1-2007) with some states following no particular standard. To reach as close as possible to net zero carbon and energy, each statewide energy code requires different levels of interventions for each code minimum building. This paper presents a collection of potential HVAC retrofits to transition each State's current code minimum buildings towards the goal of net zero to achieve a carbon free future by 2050. The study shall use a large array of code minimum criteria and climate zones covering the 48 contiguous United States to determine the most successful interventions at reducing the energy use of buildings meeting today's energy codes. Office use type has been selected for the study as they account for 18% of total buildings and close to 19% of the total commercial floorspace. A number of interventions will be applied to this use type that will vary based on current code and climate zone; however, a common theme will be electrification through the use of heat pump technology. Each intervention will be scored based on energy and carbon savings, along with level of difficulty and cost. The study not only provides a comprehensive transition guideline toward the net zero energy and carbon but also predicts the future project opportunities.*

**INTRODUCTION**

Buildings are responsible for 40% of all US primary energy use and associated carbon emissions (Wu and Skye 2018) warming caused by the concentration of such greenhouse gas emissions has led to a 1 to 3.7 degree C (1.8 to 6.6 degree F) increase in the earth temperature which continues to cause climate change (Shaikh et al. 2018). Thus, Net Zero Buildings (NEZs) are recently being seen as a promising solution to reduce human carbon footprint by 2050 (Feng et al. 2019). NEZs are often defined as buildings producing at least the amount of site energy that they consume (Sartori, Napolitano, and Voss 2012). Several renewable energy resources and envelope configurations have been investigated to achieve net zero energy goals (Wu, Skye, and Domanski 2018). However, heating, ventilation and

**Patrick Pease** is the mechanical engineering director at cove.tool, Atlanta GA. **Jayati Chhabra** and **Zahra Zolfaghari** are researcher cove.tool, Atlanta, GA.

cooling (HVAC) investigations have been less studied and are usually limited to residential buildings or certain climatic zones (Wu and Skye 2018; Wu, Skye, and Domanski 2018).

Commercial buildings contribute to 19% of the total building energy consumption ("Frequently Asked Questions (FAQs) - U.S. Energy Information Administration (EIA)" n.d.). Since the thermal comfort in these buildings such as an office has a direct impact on the productivity of the occupants, HVAC systems use energy to maintain the indoor comfort level. As one of the most energy consuming elements of buildings, HVAC systems contribute to nearly half of the total building energy consumption (Pérez-Lombard, Ortiz, and Pout 2008). Thus, energy efficient measures aiming towards improvement of the HVAC equipment can prove to be a viable solution to reach closer to net zero targets.

This study, thus, reviews the energy, cost, and operational carbon emissions associated with different HVAC interventions in different climate zones and energy codes throughout USA to propose the most energy efficient HVAC options per climate and energy code. The energy consumption for a Pacific Northwest National Laboratory (PNNL) medium office building with the baseline system along with 5 HVAC interventions has been analyzed with a preliminary energy simulation tool (cove.tool 2021) to give a better understanding of future HVAC energy efficiency requirements in different energy codes. The interventions chosen to be analyzed were based on several studies conducted in the recent years that have explored the effective ways to reduce the HVAC energy consumption. These studies ranged from incorporating more efficient equipment to enhancing the control system and operational schedule.

## METHODOLOGY

### CLIMATIC REGIONS AND CODES

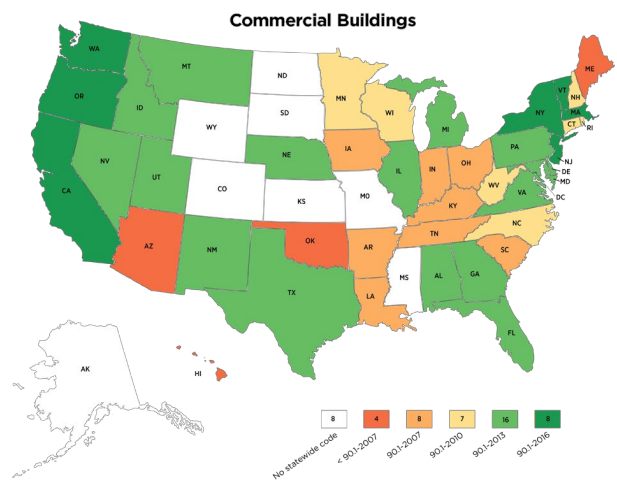

**Figure 1**     Status of stage energy code adoption in USA

According to The American Society of Heating, Refrigerating and Air-Conditioning Engineers (ASHRAE), USA is divided into 19 climate zones (ASHRAE 2019). There is no single national energy standard across the country and most of the states locally adopt ASHRAE 90.1 Energy Standard for Buildings Except Low-Rise Residential Buildings (referred as ASHRAE 90.1 in the entire paper) or IECC equivalent (IECC 2012). Energy codes define the minimum performance requirements for each state and tend to vary from ASHRAE 90.1 - 2007 to later versions of the code (**Figure 1**). Thus, based on an amalgamation of the energy codes and climatic zones (eg. ASHRAE 90.1 -2007 –

Climate zone 1A), 37 unique combinations were found.

## DEFINING BASELINE GEOMETRY AND SYSTEM

Due to ongoing growth in economy and population, the energy consumption for commercial buildings is predicted to increase. Therefore, improving the energy efficiency of commercial buildings have a significant impact on the nation's total energy consumption and carbon emission. To address that, this study focuses on the impact of HVAC system on the whole building energy use in the office buildings. To maintain consistency and narrow down the number of impactful parameters in the study such as size, geometry, WWR, etc., the PNNL prototype medium office building was used as a sample model to run different HVAC interventions and measure the savings. Also, a fixed medium density context was selected to place in each of the locations with different climate zones for reducing the impact of context on the energy results in various locations. The building is shown in Figure 2 and is made up of 3 floors, 53,600 ft² (4979 m²), an aspect ratio of 1.5 and 33% WWR.

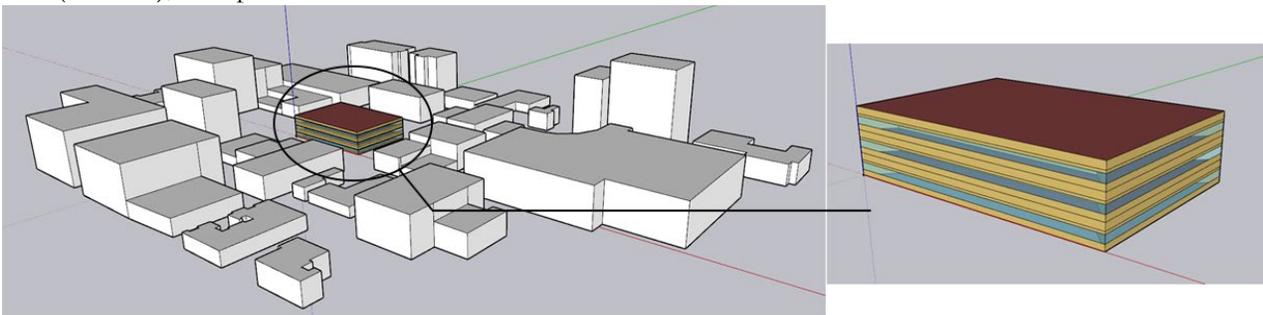

**Figure 2**   Image of model with context used

The baseline HVAC system model is assigned to reflect a typical installation for a new office building construction without any upgrades or optimizations. packaged equipment is used that would be purchased off the shelf in standard sizes for standard applications. A variable air volume (VAV) with reheat coils is the primary air system with a natural gas boiler providing heating and a packaged DX system providing cooling. The system is representative of ASHRAE 90.1 Appendix G system 5 – Packaged VAV with reheat (ASHRAE 2019). A baseline scenario is run for each of the 37 climate zone energy code pairs providing a starting point EUI, cost and carbon value. The baselines are used when calculation the percent savings of each intervention.

## METRICS

Several metrics are recorded for each simulation run that allows for comparison of the energy, cost and carbon performance of a given scenario. Energy, reported as an energy use intensity (EUI) is the primary metric of our study. This EUI is then used to calculate cost and carbon, along with savings for each. A final metric was calculated by comparing the scenario's performance to zero energy performance targets established by NBI in 2019 (Carbonnier 2019). Table 1 includes information on each metric, calculation method, and units.

**Table 1. Metrics**

| Metric | Calculation Method | Units |
|---|---|---|
| EUI | Direct from Simulation | kBtu/ft²/year (kWh/m²/year) |
| Energy Savings | Intervention energy – Baseline energy / Intervention energy | Percentage |
| Cost | Gas consumption * normalized gas rate + electric consumption * normalized electric rate | USD $ |
| Cost Savings | Intervention cost – Baseline cost / intervention cost | Percentage |
| Carbon | Gas consumption * gas emission rate + electric consumption * | $CO_2e$ |
| Carbon Savings | normalized electric emission rate Intervention carbon– Baseline carbon / Intervention carbon | Percentage |
| Distance to Zero Target | Intervention EUI - Climate Zone Net Zero Target EUI | kBtu/ft²/year (kWh/m²/year) |

Normalized values are used for cost and carbon emissions. This method was elected due to the variability of utility rates, and grid emission factors at the state level. When this variability was over laid onto the climate zone and energy code mapping, some pairs represented by states with atypically low utility costs or emissions factors skewed some results. By normalizing to standard factors for the USA, the cost and carbon factors are representative no matter the State in which a project is built. The normalized values for carbon emissions of electricity used is 0.000403 $CO_2e$/kWh taken from the EPA eGRID dataset (Environmental Protection Agency n.d.). The cost values for natural gas and electricity taken from the EIA datasets are $0.0276 per kWh gas and $0.1113 per kWh electricity (US Department of Energy n.d.; US Energy Information Administration n.d.). These values are used for all calculations of cost, carbon emissions and their savings.

## PROPOSED INTERVENTIONS

A common means of reducing energy consumption within a project is to study and implement a series of energy conservation measures (ECM) or interventions (Cleveland and Fox-Penner 2019). These can range from large impact deep energy retrofits (DER) touching everything from the envelope to HVAC, to simple lighting upgrades. While many states such as Massachusetts, New York and California are actively pursuing all means of converting their building stock to net zero carbon, other states are not strictly making such attempts. While interest and investment in net zero policies and construction is growing, a universal adoption of these policies in every state is unlikely. This potentially leaves millions of buildings without an incentive to improve (Marohl, Taub, and Lawrence 2020). Therefore, this study takes a pragmatic approach that applies to all the climate zones and states.

Assuming a lack of policy support for deep energy retrofits, this study will consider simple, low impact and low-cost interventions. Each interventions considers a single piece of HVAC equipment that can be replaced or upgraded to provide energy savings. By studying each intervention for every climate zone and minimum energy code pair in the continental United States, we aim to provide guidance for locations that do not currently have government sponsored studies.

Five interventions are considered each focusing on one of the three primary systems: heating, cooling, and controls. Table 2 summarizes each intervention with a description and details on the modeling approach.

**Table 2. Intervention Definitions**

| Number | Description | Details |
|---|---|---|
| Intervention 1 | Simple Fuel Switching | Replace natural gas boiler with an electric boiler with efficiency of 95% |
| Intervention 2 | Fuel Switching with Heat Pump | Replace natural gas boiler with an Air Source Heat Pump (ASHP) with COP of 2.05 |
| Intervention 3 | Upgrade Cooling Equipment | Replace packaged DX system with high efficiency version with COP of 4.5 |
| Intervention 4 | Building Energy Management System (BEMS) | Install and use a BEMS with advanced controls for economizer, temperature resets, prevention of simultaneous heating & cooling |
| Intervention 5 | Upgrade HVAC and lighting sensors | Add daylight sensors, occupancy sensors in all applicable spaces and optimize building schedules |

### RESULTS

For each climate zone and energy code pair a baseline and five interventions are simulated producing an EUI and all other metrics. In total 222 simulations have been run for this study providing a data set covering the continental United States. The baseline results range from a maximum EUI of 48.15 kBtu/ft²/year (151.9 kWh/m²/year) for Climate Zone 6A and ASHRAE 2010 to a minimum EUI of 24.95 kBtu/ft²/year (78.7 kWh/m²/year) for Climate zone 5C and ASHRAE 2013. After interventions are applied, the minimum EUI achieved is 22.48 kBtu/ft²/year (70.9 kWh/m²/year) for climate zone 5C and ASHRAE 2013, using intervention 5.

Looking at the entire dataset in **Figure 3** confirms several hypotheses while also pointing out some important differences between climates. As found by several previous studies intervention 2, heat pumps, has the largest potential savings at 26.6% for Climate Zone 6A ASHRAE 2010. However, it is important to note that heat pumps have low savings between 0% and 7% for climate zone 1A to 4A. This is reflective of the low heating demand of these climates and illustrates an important aspect of this study. Mainly those common conceptions of energy efficient solutions may not be applicable in all locations.

For warmer climate zones (1-4), intervention 5 (daylight sensors and control upgrade) is the best performing intervention with peak savings of 19%. This intervention focuses on reducing internal electric and heat loads which provide a two-fold savings to buildings because the electric consumption for lighting is reduced along with the cooling energy required to address it. Conversely, intervention 5 has lowest savings for the coldest climate zones (6-7) where the additional heat provides some self-heating to the building. The most consistent energy savings across all climate zones and energy codes is intervention 4 with an average of 6% savings. The lowest performing intervention is number 1 (electric boiler) with an average of only 2% over all climate zones and energy codes.

By looking at the cost savings in **Figure 4** for each intervention, different conclusions can be drawn. Immediately we see that both intervention 1 and 2 have negative cost savings which is due to the cost difference between natural gas and electricity. This is a problem for projects seeking to saving energy via fuel switching, however being stopped by the poor economic performance of such savings. Intervention 5 remains the best performing with a max of savings of 22.5% for climate zone 4A and ASHRAE 2007 and an average savings of 14% across all climate zones and energy codes. Intervention 3 and 4 are relatively flat for cost savings, averaging at only 5%.

Carbon savings in Figure 5 show a similar pattern as cost, whereas switching from natural gas to electricity is not yet a lower carbon emission solution. As such Intervention 1 show very poor performance with a minimum of -19% for climate zone 6B ASHRAE 2010. Even the efficient use of heat pumps, intervention 2 shows only a max potential carbon savings of 9% for climate zone 6A ASHRAE 2010. Comparing this to the 27% energy savings for the same pair highlights the importance of comparing all three metrics together, along with the ongoing importance of renewable energy generation and removing carbon emission of the electricity grid.

Finally, the results are tabulated against the target net zero EUI values to determine if net zero can be achieved. If an intervention brings the EUI to the target value, this infers that there is potential to generate enough energy on site (typically through solar PV) to offset the energy use. Table 3 shows that some are within 3 – 2 kBtu/ft²/year (9 - 6 kWh/m²/year) EUI of the target, however no single intervention achieves the required performance. The top 10% performing models are highlighted in green and the lowest 10% of models are highlighted in red.

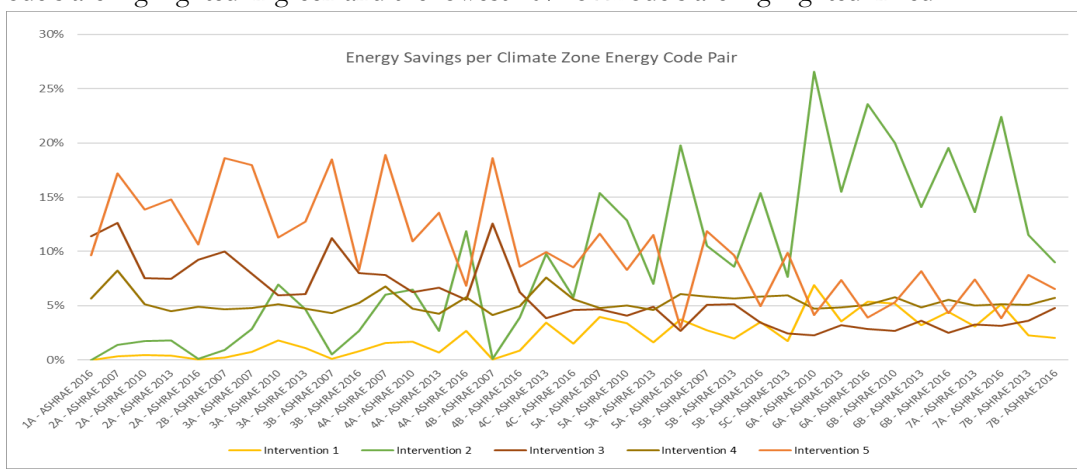

**Figure 3**    Energy Savings per Climate Zone Energy Code Pair

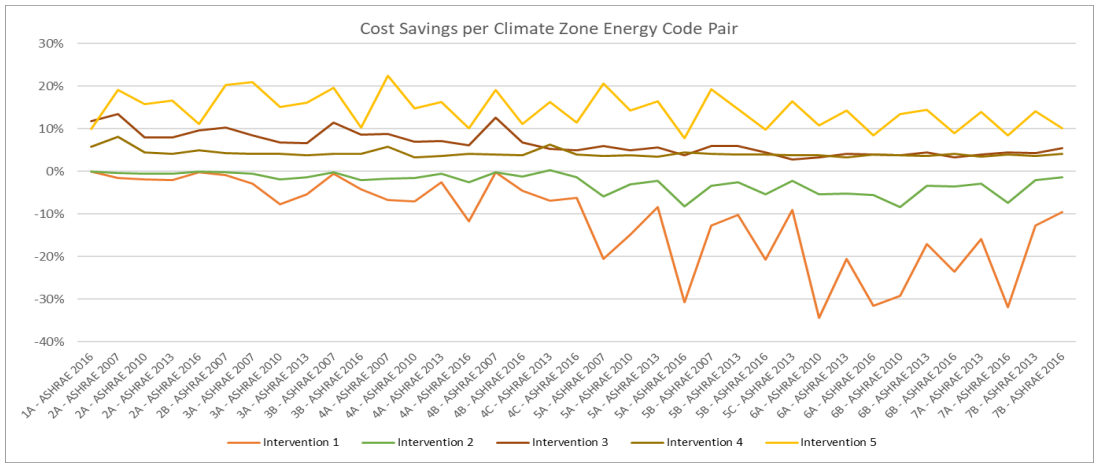

**Figure 4**    Cost Savings per Climate Zone Energy Code Pair

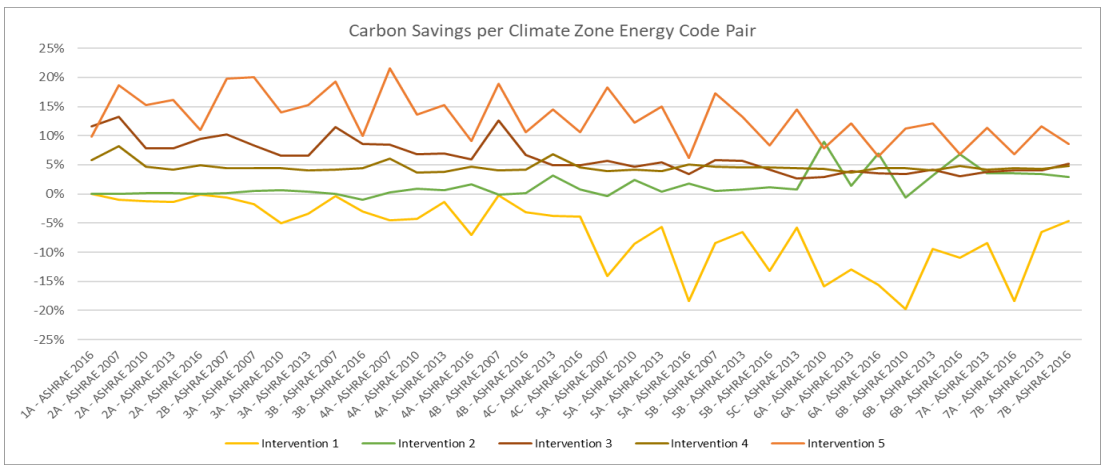

**Figure 5**    Carbon Savings per Climate Zone Energy Code Pair

### Table 3. Intervention EUI performance and Distance to Net Zero Target

|  | Intervention 1 | Intervention 2 | Intervention 3 | Intervention 4 | Intervention 5 |
|---|---|---|---|---|---|
| 1A - ASHRAE 2016 | 9.2 (29) | 9.2 (29) | 5.4 (17.1) | 7.3 (23.1) | 6 (18.9) |
| 2A - ASHRAE 2007 | 8.5 (26.8) | 8.2 (25.8) | 4.6 (14.6) | 6 (18.9) | 3.2 (10) |
| 2A - ASHRAE 2010 | 8.6 (27.2) | 8.2 (25.9) | 6.4 (20.1) | 7.1 (22.5) | 4.4 (13.8) |
| 2A - ASHRAE 2013 | 5.9 (18.7) | 5.5 (17.4) | 3.9 (12.2) | 4.8 (15) | 1.8 (5.6) |
| 2A - ASHRAE 2016 | 6.1 (19.2) | 6.1 (19.1) | 3.4 (10.8) | 4.7 (14.7) | 3 (9.5) |
| 2B - ASHRAE 2007 | 14.6 (46) | 14.3 (45.1) | 10.9 (34.4) | 12.9 (40.7) | 7.7 (24.1) |
| 3A - ASHRAE 2007 | 12.5 (39.4) | 11.7 (36.9) | 9.9 (31.2) | 11 (34.8) | 6.3 (19.9) |
| 3A - ASHRAE 2010 | 9.6 (30.4) | 7.9 (25) | 8.3 (26.1) | 8.5 (26.9) | 6.5 (20.5) |
| 3A - ASHRAE 2013 | 5.6 (17.5) | 4.5 (14.2) | 4.1 (13) | 4.5 (14.2) | 2.2 (6.9) |
| 3B - ASHRAE 2007 | 18.7 (58.9) | 18.5 (58.4) | 14.3 (45) | 17 (53.6) | 11.4 (35.9) |
| 3B - ASHRAE 2016 | 8.2 (25.9) | 7.6 (24.1) | 6.1 (19.1) | 6.9 (21.7) | 6 (18.9) |
| 4A - ASHRAE 2007 | 16 (50.5) | 14.3 (45) | 13.6 (42.8) | 14 (44.1) | 9.3 (29.3) |
| 4A - ASHRAE 2010 | 10.9 (34.4) | 9.3 (29.3) | 9.4 (29.6) | 9.9 (31.2) | 7.8 (24.6) |
| 4A - ASHRAE 2013 | 6.1 (19.1) | 5.5 (17.3) | 4.4 (13.8) | 5.1 (15.9) | 2.4 (7.6) |
| 4A - ASHRAE 2016 | 7.1 (22.5) | 4.4 (13.8) | 6.3 (19.8) | 6.2 (19.6) | 5.9 (18.6) |
| 4B - ASHRAE 2007 | 22.4 (70.5) | 22.3 (70.4) | 17 (53.8) | 20.6 (65) | 14.5 (45.7) |
| 4B - ASHRAE 2016 | 6.8 (21.5) | 6 (18.9) | 5.4 (16.9) | 5.7 (18) | 4.7 (14.9) |
| 4C - ASHRAE 2013 | 4.6 (14.4) | 3 (9.4) | 4.5 (14.1) | 3.5 (11.1) | 2.9 (9.2) |
| 4C - ASHRAE 2016 | 5.1 (16.1) | 4 (12.7) | 4.3 (13.6) | 4.1 (12.8) | 3.3 (10.5) |
| 5A - ASHRAE 2007 | 16.7 (52.6) | 11.8 (37.4) | 16.4 (51.6) | 16.3 (51.5) | 13.4 (42.4) |
| 5A - ASHRAE 2010 | 6.8 (21.5) | 3.8 (11.9) | 6.6 (20.8) | 6.3 (19.8) | 5.3 (16.6) |
| 5A - ASHRAE 2013 | 4.2 (13.4) | 2.7 (8.5) | 3.3 (10.4) | 3.4 (10.6) | 1.4 (4.4) |
| 5A - ASHRAE 2016 | 9 (28.3) | 3.5 (10.9) | 9.3 (29.4) | 8.2 (25.7) | 9.2 (29.1) |
| 5B - ASHRAE 2007 | 9.7 (30.7) | 7.3 (22.9) | 9 (28.4) | 8.8 (27.6) | 6.8 (21.6) |
| 5B - ASHRAE 2013 | 7.9 (25) | 6 (18.9) | 7 (22.1) | 6.8 (21.6) | 5.7 (17.9) |
| 5B - ASHRAE 2016 | 9.4 (29.5) | 5.6 (17.7) | 9.4 (29.6) | 8.6 (27.2) | 8.9 (28.1) |
| 5C - ASHRAE 2013 | 4.5 (14.2) | 3 (9.6) | 4.3 (13.7) | 3.5 (10.9) | 2.5 (7.8) |
| 6A - ASHRAE 2010 | 19.8 (62.6) | 10.4 (32.7) | 22.1 (69.6) | 20.9 (65.8) | 21.2 (66.8) |
| 6A - ASHRAE 2013 | 6.7 (21.1) | 2.8 (8.7) | 6.8 (21.5) | 6.3 (19.7) | 5.4 (17.2) |
| 6A - ASHRAE 2016 | 12.4 (39.1) | 5.2 (16.5) | 13.4 (42.3) | 12.5 (39.5) | 13 (40.9) |
| 6B - ASHRAE 2010 | 14.3 (45.1) | 8.5 (26.7) | 15.3 (48.2) | 14.1 (44.4) | 14.3 (45) |
| 6B - ASHRAE 2013 | 8.2 (25.9) | 4.7 (14.8) | 8.1 (25.5) | 7.7 (24.3) | 6.6 (20.9) |
| 6B - ASHRAE 2016 | 9.1 (28.6) | 4 (12.6) | 9.7 (30.6) | 8.7 (27.4) | 9.1 (28.7) |
| 7A - ASHRAE 2013 | 8 (25.3) | 4.8 (15) | 8 (25.1) | 7.4 (23.4) | 6.7 (21) |
| 7A - ASHRAE 2016 | 14.7 (46.4) | 8 (25.4) | 15.5 (48.8) | 14.7 (46.4) | 15.2 (48) |
| 7B - ASHRAE 2013 | 7.3 (23.1) | 4.6 (14.4) | 6.9 (21.9) | 6.5 (20.5) | 5.7 (17.9) |
| 7B - ASHRAE 2016 | 5.8 (18.4) | 3.8 (12.1) | 5 (15.9) | 4.8 (15.1) | 4.5 (14.3) |

**CONCLUSION**

By looking at the entire data space for energy, cost, and carbon savings against each climate zone and energy code pair a clear picture of best practice in each region of the USA emerges. First, we can consider heat pumps and their correct application and challenges. Intervention 2 (heat pumps) was shown to have limited energy savings for any energy codes in climate zone 1A to 4A, this means projects do not need to consider this option when targeting energy savings. The application of intervention 2 in climate zone 5A-7B produce good energy savings; however, the heat pumps result in slightly more cost and carbon emissions in these same climate zones. This highlights the need for parallel improvements to electricity cost and carbon emission intensity in order to fully see the benefits of heat pump installations.

Intervention 5, upgraded lighting and HVAC controls, shows great potential for energy, cost and carbon savings. This intervention is unique as it reduces unregulated lighting load for double benefit in cooling dominated climates. Also, in heating dominated climates the additional heat gain from lights can contribute moderately to self-heating. Upgrading of control system with sensors etc. is relatively low first cost and low disruption to a building's occupant, which makes it a great option for buildings looking for an easy path towards more sustainability.

The results of this study can be used in two ways. First, for project teams designing buildings without a budget for energy modeling or system analysis, the results can be used as a guide to pick the single best improvement to their design given the climate zone and energy code they are working in. The results can also be used to make decisions for current projects based on what future projects will be like in 10, 20, or 30 years. Given that a large portion of buildings built today are not focused on sustainability, they are built to meant only the minimum energy codes in their location. By using the energy code and climate zone mapping of this study, firms can identify what interventions will be most common as these current buildings eventually move to reduce their energy use. For example, for a firm working in Louisiana or Mississippi where the current energy codes are 2007 and 2010 respectively, it can be taken from this study that control system upgrades and sensor installations will be important in the future. Therefore, firms could take steps now to ready themselves to be able to capture that market.

Finally, this study shows the wide differences in climate of the continental United States results in there being no single solution that is best everywhere. In discussing energy savings and transitioning of buildings to net zero energy, location is an important aspect to consider. Heat pumps, while critical in colder regions, are not as helpful in warmer regions where their savings will almost never payback the initial installation costs.